\documentclass[showpacs]{jetpl}
\twocolumn
\usepackage[cp1251]{inputenc} 
\usepackage[english,russian]{babel}
\usepackage{caption}

\lat

\begin{document}

\title{Dependence of five and six-loop estimated QCD corrections to the relation between pole and running masses of heavy quarks on the number of light flavours}

\rtitle{Dependence of five and six-loop estimated QCD corrections \ldots}


\author{A.\,L.\,Kataev$^{\,ab}$\/\thanks[1]{e-mail: kataev@ms2.inr.ac.ru},
V.\,S.\,Molokoedov$^{\,abc}$\/\thanks[2]{e-mail: viktor\_molokoedov@mail.ru}}

\rauthor{A.\,L.\,Kataev, V.\,S.\,Molokoedov}

\sodauthor{Kataev, Molokoedov}

\address{$^a$Institute for Nuclear
Research RAS, 117312 Moscow, Russia\\~\\
$^b$Moscow Institute of Physics and Technology, 141700 Dolgoprudny, Russia\\~\\
$^c$L.\,D.\,Landau Institute for Theoretical Physics RAS,
142432 Chernogolovka, Russia}

\dates{8 October 2018}{*}

\abstract{In this paper various theoretical approaches are  used to define the dependence of the
estimated  $\mathcal{O}(\alpha^5_s)$ and $\mathcal{O}(\alpha^6_s)$-corrections to the QCD relation between pole and $\rm{\overline{MS}}$ running masses of heavy quarks
on the number of light flavours. It is found that  recently studied  asymptotic formula for the coefficients of this relation, based on the infared-renormalon method, does not reproduce sign-alternating structure in the flavour-dependence of the five and six-loop corrections, which holds in three other used by us approaches.}


\maketitle

\section{Inroduction}
The  masses of charm, bottom and top-quarks are the important parameters not only of QCD, but of the Standard Model of particle physics and of its various extensions. However, due to the phenomenon of confinement  quark masses can not be measured directly. Moreover, within the perurbative QFT masses depend not only on the energy scale $\mu$, but also on the concrete renormalization scheme. Therefore there are several theoretical definitions of heavy quarks masses. 
The widespread mass notions are the running $\overline{m}_q(\mu^2)$ and pole masses $M_q$, determined  within the  $\rm{\overline{MS}}$-scheme and the on-shell (OS) renormalization scheme correspondingly.  The relation between these  masses (we will call it briefly  $\rm{\overline{MS}}$-on-shell relation)
was considered in the number of works at one- \cite{Tarrach:1980up}, two- \cite{Gray:1990yh, Avdeev:1997sz, Fleischer:1998dw} and three-loop \cite{Melnikov:2000qh, Chetyrkin:1999qi} level and has the following form
\begin{equation}
\label{t^M_n}
M_q=\overline{m}_q(\overline{m}^2_q)\sum\limits_{n\geq 0} t_n a^n_s(\overline{m}^2_q),
\end{equation}
where $\mu^2=\overline{m}^2_q$ and $a_s(\mu^2)=\alpha_s(\mu^2)/\pi$ in the $\rm{\overline{MS}}$-scheme. For case of the $SU(3)$ color gauge group the numerical results of these analytical calculations read:
\begin{align}
\label{t0-3}
t_0&=1, ~~~ t_1=4/3, ~~~ t_2=-1.0414n_l+13.443, \\ \nonumber
t_3&=0.6527n^2_l-26.655n_l+190.60.
\end{align}
Corrections of the second and third order of perturbation theory (PT) depend on $n_l$. In this work we consider the case of one massive and $n_l$ massless quarks, i.e. the number of active flavours  $n_f=n_l+1$. This approximation is really very good if we ignore the corrections associated with the inclusion of quark masses in the internal fermionic inserts which renormalize the quark two-point Green function \cite{Bekavac:2007tk}.  As it follows from eqs.(\ref{t0-3}) the structure of expressions for coefficients $t_n$ is sign-alternating in $n_l$. Study of this problem at the fourth-loop level confirms this rule. However, at present the analytical form of $t_4$-coefficient is not yet known. But the leading and sub-leading in powers of $n_l$ terms were obtained analytically \cite{Lee:2013sx} and the first of them  is in agreement with the results of the renormalon-based analysis \cite{Ball:1995ni}. Further, in the work of \cite{Marquard:2015qpa} values for full expression of $t_4$-coefficient were evaluated numerically with the identical uncertainties for  $n_l=3, 4$ and $5$.  Taking into account the results of these computations and applying the mathematical least squares method (LSM) to solve the overdetermined system of three linear equations with two unknowns (the linearly dependent on $n_l$ and irrespective on $n_l$ terms in expansion of $t_4$) one can obtain not only their numerical values, but also evaluate their theoretical inaccuracies \cite{Kataev:2015gvt}. Recently results \cite{Marquard:2015qpa}  were improved   in  Ref.\cite{Marquard:2016dcn} by  presenting  numerical values for $t_4$-term at $0\leq n_l\leq 20$ with much higher $n_l$-dependent  numerical  accuracy. This circumstance contributed to the reconsideration of the LSM-calculations \cite{Kataev:2015gvt} in the work \cite{Kataev:2018gle}. The updated results \cite{Kataev:2018gle} of flavour dependence for the $\mathcal{O}(a^4_s)$-term in the $\rm{\overline{MS}}$-on-shell mass relation of Eq.(\ref{t^M_n}) read:
\begin{align}
\label{t4}
t_4&=-0.6781n^3_l+43.396n^2_l \\ \nonumber
&+(-745.72\pm 0.15)n_l+3567.60\pm 1.34.
\end{align}
Analysis of the relation between pole and running masses of heavy flavours at the fourth-loop level points out that the corresponding asymptotic PT series for the charm-quark diverges from the second (or the third) order of PT. In case of $b$-quark all high-order corrections decrease up to four-loop level. However, the $\mathcal{O}(a^4_s)$-contribution becomes comparable in magnitude with the $\mathcal{O}(a^3_s)$-term. Therefore for an exact answer to the question of finding an order up to which the truncated $\rm{\overline{MS}}$-on-shell series for $b$-quark  can be used in theoretical studies within the PT, it is necessary to define the fifth-order corrections to this relation at least. For $t$-quark this task is even more relevant.
Indeed, all high-order contributions
decrease monotonically and rather quickly up to four-loop level. In addition, due to the large mass of $t$-quark the magnitudes of these corrections are noticeably larger than for the  case of $b$-quark and therefore taking into account the effects of high orders of PT for $t$-quark plays important role in definition of theoretical uncertainties for mass values, extracted from the experimental data at
Tevatron and LHC. In fact, the theoretical errors of the masses of heavy quarks are estimated by the order of magnitude of the last included PT correction in the ratio between the pole and running masses of heavy flavours. Note, that the mentioned asymptotic nature of the $\rm{\overline{MS}}$-on-shell mass relation is related to the existence of the leading 
infrared renormalon (IRR) singularity in the Borel image of  Eq.(\ref{t^M_n}) \cite{Bigi:1994em, Beneke:1994sw},
which does coefficients of the perturbative series for this relation factorially growing. Thus, it is important from theoretical and phenomenological point of view to fix the number of order of PT from which the corresponding series for the $\rm{\overline{MS}}$-on-shell relation starts to manifest the asymptotic character. Let's move on to the study of this problem.
\section{Estimates based on resummation of the renormalon chains}
In this section we use the results of work \cite{Ball:1995ni}, where corrections to the relation between pole and running heavy quark masses were estimated using the calculations of the leading on $n_l$ contributions
from consideration of the chain of fermion loops (FL) into the gluon propagator, renormalizing two-point quark Green function, supplemented by the procedure of naive nonabelianization, which in the normalization, used in this work, is  equivalent to the substitution $n_l\rightarrow -6\beta_0$, where $\beta_0=11/4-(n_l+1)/6$ is the first scheme-independent coefficient of the QCD renormalization-group  $\beta$-function.  We use the results of these computations, shifting the normalization point from the pole mass to the $\rm{\overline{MS}}$-scheme running mass of heavy flavour. In case of applying this method to the estimation of the four-loop correction to the $\rm{\overline{MS}}$-on-shell mass relation we obtain following expression
\begin{equation}
\label{t4FL}
t^{FL}_4=-0.678n^3_l+30.66n^2_l-435.5n_l+2145,
\end{equation}
which is in satisfactory agreement with (\ref{t4}).
For coefficients $t_5$ and $t_6$  this renormalon-chain   procedure gives the following predictions:
\begin{align}
\label{t5FL}
t^{FL}_5&=0.9n^4_l-56n^3_l+1256n^2_l-12383n_l+47721, \\
\label{t6FL}
t^{FL}_6&=-1.5n^5_l+120n^4_l-3779n^3_l \\ \nonumber &+58846n^2_l-460910n_l+1468466.
\end{align}
Let us stress that equations (\ref{t4FL}),  (\ref{t5FL}) and (\ref{t6FL}) contain the exact numerical  expressions of leading in powers of $n_l$ contributions. Note also that this procedure reproduces explicitly the sign-alternating structure of $t^{FL}_4$, $t^{FL}_5$ and $t^{FL}_6$-terms, which manifests itself at two, three and four-loop levels.
\section{Application of the effective charges method}
\subsection*{Estimates in the Minkowskian region}
Consider the analogue of the equation (\ref{t^M_n}), namely the quantity $T(s)$, defined in the Minkowskian time-like region with $\mu^2=s$ as:
\begin{equation}
\label{T(s)}
T(s)=\overline{m}_q(s)\sum\limits_{n\geq 0} t_n a^n_s(s),
\end{equation}
where coefficients $t_n$ at $0\leq n \leq 4$ coincide with the calculated analytically in Eq.(\ref{t0-3}) and with the semi-analytical result (\ref{t4}). Further we use the method for estimations of the high-order PT QCD corrections to the  quantity  $T(s)$, based on the concept of the effective charges (ECH) \cite{Grunberg:1982fw} and developed in the  works \cite{Kataev:1995vh, Chetyrkin:1997wm} approach. At the first stage we define the effective charge $a^{eff}_s(s)$ for the quantity $T(s)/\overline{m}_q(s)$, viz
\begin{equation}
a^{eff}_s(s)=a_s(s)+\sum\limits_{n\geq 2} \tau_n a^n_s(s)
\end{equation}
with $\tau_n=t_n/t_1$. After this we can introduce the corresponding ECH $\beta$-function, which is responsible for the evolution of the $a^{eff}_s(s)$ coupling constant by the following way:
\begin{equation}
\beta^{eff}(a^{eff}_s)=-\sum\limits_{n\geq 0} \beta^{eff}_n (a^{eff}_s)^{n+2}.
\end{equation}
Scheme-independent coefficients $\beta^{eff}_n$ for $n=4,5$ are related to the $\rm{\overline{MS}}$ coefficients $\beta_n$ by the following  relations \cite{Kataev:1995vh}:
\begin{eqnarray}
\label{betaef0-5}
\beta^{eff}_4&=&\beta_4-3\tau_2\beta_3+(4\tau^2_2-\tau_3)\beta_2+(\tau_4-2\tau_2\tau_3)\beta_1 \\ \nonumber
&+&(3\tau_5-12\tau_2\tau_4-5\tau^2_3+28\tau^2_2\tau_3-14\tau^4_2)\beta_0, \\ \nonumber
\beta^{eff}_5&=&\beta_5-4\tau_2\beta_4+(8\tau^2_2-2\tau_3)\beta_3+(4\tau_2\tau_3-8\tau^3_2)\beta_2 \\ \nonumber
&+&(2\tau_5-8\tau_2\tau_4+16\tau^2_2\tau_3-3\tau^2_3-6\tau^4_2)\beta_1 \\ \nonumber
&+&(4\tau_6-20\tau_2\tau_5-16\tau_3\tau_4+48\tau_2\tau^2_3
-120\tau^3_2\tau_3 \\ \nonumber
&+&56\tau^2_2\tau_4+48\tau^5_2)\beta_0.
\end{eqnarray} 
The ansatz $\beta^{eff}_n=\beta_n$ allows to estimate the numerical value of  
$\tau_{n+1}$-coefficient.  
This procedure was applied in \cite{Chetyrkin:1997wm, Kataev:2010zh}, where corrections $t_3$ and $t_4$ to the relation 
(\ref{t^M_n}) for $c$, $b$ and $t$-quarks were obtained. 
These estimates turned out to be  
in rather good agreement with the results of the explicit $\mathcal{O}(a^3_s)$ and $\mathcal{O}(a^4_s)$ diagram-by-diagram calculations (\ref{t0-3}) and (\ref{t4}). Therefore we expect that approximations $\beta^{eff}_4=\beta_4$ and $\beta^{eff}_5=\beta_5$ will lead to reasonable estimates
for five- and six-loop corrections to the $\rm{\overline{MS}}$-on-shell mass relation as well
(we denote them $t^{ECH-M}_5$ and $t^{ECH-M}_6$). Considering now the cases with fixed numbers of massless flavours $3\leq n_l\leq 8$ (which do not change the sign of the $\beta_0$-coefficient and therefore do not contradict the property of asymptotic freedom of QCD in the leading order approximation) we predict unequivocally values of all coefficients in the flavour dependence of the five- and six-loop corrections. Within the ECH-motivated approach, applied directly in the time-like region, the expressions 
for $\mathcal{O}(a^5_s)$ and $\mathcal{O}(a^6_s)$-corrections
have the following form:
\begin{align}
\label{t_5-ECH-M}
t^{ECH-M}_5&=1.2n^4_l-77n^3_l \\ \nonumber
&+1959n^2_l-20445n_l+72557, \\
\label{t_6-ECH-M}
t^{ECH-M}_6&=-2.2n^5_l+148n^4_l-4561n^3_l \\ \nonumber &+71653n^2_l-538498n_l+1519440.
\end{align}
One can see that they are in quite acceptable accordance with the results (\ref{t5FL}) and (\ref{t6FL}), predicted by means of the renormalon-chain calculations. Both these approaches indicate the sign-alternating structure in expansion of five- and six-loop corrections in powers of $n_l$.
\subsection*{Estimates with transition from the Euclidean to Minkowskian region}
In practice most of PT calculations are performed in the Euclidean region, whereas the physical characteristics of  processes, which may be measured at colliders, are described  by quantities in the time-like regions of energies. In order to establish matching between them it is necessary to take into account the effects, related to the transition from the  Euclidean to Minkowskian  space. As was demonstrated  in Refs.\cite{Kataev:1995vh, Chetyrkin:1997wm}  it is more theoretically substantiated to apply the ECH-motivated approach to physical quantities, defined in the Euclidean region. After that it is necessary to evaluate the effects of analytical continuation and get the expression for the quantities, measured in the Minkowski region. We consider the K\"allen-Lehman type spectral representation of the Euclidean quantity $F(Q^2)$ \cite{Chetyrkin:1997wm}  
\begin{equation}
\label{F(Q^2)}
F(Q^2)=\int\limits_0^{\infty}ds \frac{Q^2T(s)}{(s+Q^2)^2}=\overline{m}_q(Q^2)\sum_{n\geq 0} f_n a^n_s(Q^2)
\end{equation}
with Minkowskian function $T(s)$, determined in Eq.(\ref{T(s)}). Taking into account the scale dependence of the $\rm{\overline{MS}}$-scheme coupling constant $a_s(s)$ and the running mass $\overline{m}_q(s)$, carrying out the integration in Eq.(\ref{F(Q^2)}) and setting $\mu^2=Q^2$, we obtain the relations between the coefficients $t_n$ and  $f_n$ of the PT series in the Minkowskian and Euclidean regions 
\begin{equation}
\label{relation}
f_n=t_n+\Delta_n,
\end{equation}
where additional contributions $\Delta_n$ are effects of the analytical continuation  and contain terms, proportional to powers of $\pi^2$ and powers of the $\beta$-function coefficients and anomalous mass dimension $\gamma_m$ in the $\rm{\overline{MS}}$-scheme. The explicit six-loop expressions for $\Delta_n$-terms 
are given in \cite{Kataev:2018gle} for case of the $SU(N_c)$ colour gauge group. For particular case of the $SU(3)$ group we have the following 
numerical expressions of these contributions:
\begin{align}
\label{C2}
\Delta_0&=0, ~~~~~~~~ \Delta_1=0, \\ \nonumber
\Delta_2&=5.8943-0.27416n_l, \\ \nonumber
\Delta_3&=105.622-10.0448n_l+0.19800n^2_l, \\
 \nonumber
\Delta_4&=2272.00-403.949n_l+20.6767n^2_l-0.31590n^3_l, \\ \nonumber
\Delta_5&=56304.64-13767.273n_l+1137.1779n^2_l \\ \nonumber 
&-37.74529n^3_l+0.42752n^4_l,  \\
 \nonumber
\Delta_6&=1633115.6\pm 347.7+(-518511.69\pm 56.72)n_l \\ \nonumber
&+(61128.167\pm 4.779)n^2_l+(-3345.082\pm 0.137)n^3_l \\ \nonumber
&+85.3794n^4_l-0.81845n^5_l.
\end{align}
The expressions of Eqs.(\ref{C2}) demonstrate that the analytical continuation contributions are not negligible and increase significantly with the growth of the order $n$ of PT. The substantial difference between method, described in this subsection, and approach of the direct application of the ECH-motivated procedure in the Minkowski region, lies in the  construction of an effective charge not for the Minkowskian quantity $T(s)/\overline{m}_q(s)$, but for the Euclidean associated function $F(Q^2)/\overline{m}_q(Q^2)$, defined in (\ref{F(Q^2)}). This change is equivalent to the replacement $\tau_n\rightarrow f_n/f_1$ in eqs.(\ref{betaef0-5}). Applying the reasoning similar to the ones, used in the previous subsection, namely equating the coefficients of the ECH $\beta$-function, defined in the Euclidean region, to their $\rm{\overline{MS}}$-scheme analogues, we get estimates of the five- and six-loop coefficients of the Euclidean coefficients $f_5$ and $f_6$ at fixed $n_l$ in the region  $3\leq n_l\leq 8$. Reproducing now their explicit expansion in powers of $n_l$  by solving the systems of the corresponding equations and taking into account the relation (\ref{relation}),  we obtain the following   estimates of the $n_l$-dependent expressions of the Minkowskian terms $t_5$ and $t_6$ (we denote them $t^{ECH-E-M}_5$ and $t^{ECH-E-M}_6$):
\begin{align}
\label{t_5-ECH-E-M}
t^{ECH-E-M}_5&=2.5n^4_l-136n^3_l \\ \nonumber
&+2912n^2_l-26976n_l+86620, \\
\label{t_6-ECH-E-M}
t^{ECH-E-M}_6&=-4.9n^5_l+352n^4_l-9708n^3_l \\ \nonumber
 &+131176n^2_l-855342n_l+2096737.
\end{align}
Like in two previous considered by us cases, the ECH-motivated approach, supplemented by taking into account  the effects of analytical continuation, respects the property of sign-alternating structure of $t_n$-corrections in the $\mathcal{O}(a^5_s)$ and $\mathcal{O}(a^6_s)$ approximations as well. Note that numerical values of $\Delta_5$ and $\Delta_6$-contributions  are comparable in magnitude with values of $t_5$ and $t_6$-terms. Herewith the values of corresponding coefficients in $n_l$-expansion of $t^{ECH-M}_{5,6}$ and $t^{ECH-E-M}_{5,6}$-contributions differ in insignificant factors from 1.5 to 2 (see expressions (\ref{t_5-ECH-M}), (\ref{t_6-ECH-M}) and (\ref{t_5-ECH-E-M}), (\ref{t_6-ECH-E-M})).
\section{The asymptotic renormalon studies}
As the next estimation method of the five- and six-loop contributions to the $\rm{\overline{MS}}$-on-shell heavy quark mass relation we consider the procedure, based on the  asymptotic formula, derived in Refs.\cite{Beneke:1994rs, Beneke:1998ui} and corrected a bit in Ref.\cite{Pineda:2001zq}. It is based on the IRR renormalon dominance
hypothesis, which appeared after studying the infrared singularities of the Borel image of the PT series for the relation between the pole and running masses of quarks \cite{Bigi:1994em, Beneke:1994sw} and goes beyond  large $\beta_0$-expansion. The general expression for this formula  is written as
\begin{align}
\label{renormalon-dom}
t^{r-n}_n\xrightarrow{n\rightarrow\infty}\pi N_m (2\beta_0)^{n-1}\frac{\Gamma(n+b)}{\Gamma(1+b)}\bigg(1+
\sum\limits_{k=1}^3\Omega_k\bigg)
\end{align}
where $\Gamma(x)$ is the Euler Gamma-function, $b=\beta_1/(2\beta^2_0)$ and expressions for $\Omega_k$-contributions, containing terms suppressed by $(1/n)^k$-corrections, are given in \cite{Pineda:2001zq, Beneke:2016cbu}. The normalization factor $N_m$ in Eq.(\ref{renormalon-dom}) depends on $n_l$ and on $n$. Possible ways to fix the values of the $N_m$-factor in concrete order of PT are presented in the works \cite{Pineda:2001zq, Campanario:2003ix, Ayala:2014yxa}. In order to estimate a magnitude of the $\mathcal{O}(a^5_s)$ and $\mathcal{O}(a^6_s)$ corrections to the $\rm{\overline{MS}}$-on-shell heavy quark mass relation within asymptotic formula (\ref{renormalon-dom}) we use numerical $\it{four-loop}$ results for $N_m$-factor, obtained for interval $3\leq n_l\leq 8$ \cite{Beneke:2016cbu}.
The possibility of such approximation follows from rather weak dependence of $N_m$-factor on the order of PT, beginning with three-loop level. Naturally, this dependence is not negligible, but for our goals to estimate the $\mathcal{O}(a^5_s)$ and $\mathcal{O}(a^6_s)$ corrections it will  be quite suitable. However, we are faced with one unexpected fact: unlike the supported by large $\beta_0$-expansion approach \cite{Ball:1995ni} and the both realizations of the ECH-motivated method \cite{Kataev:2018gle}, the application of the IRR-based formula with the $\mathcal{O}(a^4_s)$ approximation of $N_m$-factor does not reproduce the sign-alternating $n_l$-dependence of the corresponding five- and six-loop expressions for $t_5$ and $t_6$-coefficients:
\begin{align}
\label{t5ren}
t^{r-n}_5&=-22n^4_l+416n^3_l \\ \nonumber
&-1669n^2_l-11116n_l+72972, \\ 
\label{t6rn}
t^{r-n}_6&=99n^5_l-2903n^4_l+30109n^3_l \\ \nonumber
&-99563n^2_l-305378n_l+2040263.
\end{align}
This may indicate at least two circumstances:
either our assumption to use the four-loop values of $N_m $-factor in $\mathcal{O}(a^5_s)$ and $\mathcal{O}(a^6_s)$-approximations is not valid or it is necessary to take into account in the IRR-based formula additional sources of
uncertainties, say contributes of $(1/n)^4$-corrections or UV-renormalon effects \cite{Beneke:1998ui, Broadhurst:2000yc}. 
\section{Numerical results}
To summarise discussions, described in the previous sections, we present the numerical results for five- and six-loop coefficients $t_5$, $t_6$ at $3\leq n_l\leq 8$, predicted within the four considered by us estimated procedures.

From the results of Table it follows that for physical values
$n_l=3,4,5$ predictions, made within the renormalon asymptotic formula (\ref{renormalon-dom}), are slightly higher than the estimates obtained by resummation of the inserts of fermion loops and two variants of the ECH-method.
It is also worth to stress that the estimates for the $t_5$-coefficient, derived by both ECH approaches, are
in better agreement with the results, obtained in the process of the $b$-quark mass determination from the global fits of quark-antiquark bound states \cite{Mateu:2017hlz}, than with ones, extracted from the IRR-based asymptotic formula of Eq.(\ref{renormalon-dom}). Unlike the FL and IRR-based techniques both ECH-approaches indicate the negative values of $t_5$ and $t_6$-corrections for $n_l=7, 8$, which are important for their sign-alternating behavior.  

\begin{table}[h!]
\footnotesize
\begin{tabular*}{\columnwidth}{@{\extracolsep{\fill}}ccccc}
\hline 
~~$n_l$~~ & $t^{FL}_5$  & $t^{ECH-M}_5$ & $t^{ECH-E-M}_5$ & ~~~$t^{r-n}_5$~~~ \\
\hline 
3 & 20432 & 26871 & 28435 & 34048  \\
\hline
4 & 14924 & 17499 & 17255 & 22781  \\
\hline
5 & 10757 & 10427 & 9122  & 13882  \\
\hline
6 & 7693 & 5320 &  3490  & 7466  \\
\hline
7 & 5515 & 1871 &  -127  & 3119  \\
\hline
8 & 4027 & -196 &  -2153 & 344 \\
\hline\hline
~~$n_l$~~ & $t^{FL}_6$  & $t^{ECH-M}_6$ & $t^{ECH-E-M}_6$ & ~~~$t^{r-n}_6$~~~ \\
\hline
3 & 522713 & 437146 & 476522 & 829993  \\
\hline
4 & 353810 & 255692 & 238025 & 511245  \\
\hline
5 & 233282 & 133960 & 90739 & 283902  \\
\hline
6 & 149601 & 57920 &  8412 & 137256  \\
\hline
7 & 93225 & 15798 &  -29701 & 50520  \\
\hline
8 & 56410 & -2184 & -39432 &  4747  \\
\hline
\end{tabular*}
\captionsetup{size=footnotesize}
\caption*{Table: Estimates for $t_5$ and $t_6$-terms by four various methods}
\end{table}

Now consider a magnitude of the high-order QCD corrections to the $\rm{\overline{MS}}$-on-shell mass relation of real heavy quarks. We have already discussed above that in the case of the charm-quark the asymptotic structure of the $\rm{\overline{MS}}$-on-shell mass relation reveals itself from the second order of PT. Therefore  we are interested in the magnitude of corrections of the high order PT  for $b$ and $t$-quarks only. We fix the values of the running masses of these flavours and coupling constants as in work \cite{Kataev:2018gle}, namely   $\overline{m}_b(\overline{m}^2_b)=4.180 \; \rm{GeV}$, $\overline{m}_t(\overline{m}^2_t)=164.3 \; \rm{GeV}$, $\alpha_s(\overline{m}^2_b)=0.2256$,  $\alpha_s(\overline{m}^2_t)=0.1085$. Taking into account the  known results of direct diagram calculations and using the data from Table, we arrive at the following expressions obtained by four considered by us methods:
\begin{subequations}
\begin{align}
\label{Mb}
\frac{M_b}{1~\rm{GeV}}&\approx 4.180+0.400+0.200+0.146+0.137 \\ \nonumber
&+\left\{
\begin{aligned}
&0.119+0.203~~ - ~~\text{FL}; \\
&0.140+0.147~~ - ~~\text{ECH-M};\\ 
&0.137+0.137~~ - ~~\text{ECH-E-M};\\
&0.182+0.293~~ - ~~\text{IRR};\\
\end{aligned}
\right. 
\end{align}
\begin{align}
\label{Mt}
\frac{M_t}{1~\rm{GeV}}&\approx  164.3+7.566+1.614+0.498+0.196 \\ \nonumber
&+\left\{
\begin{aligned}
&0.087+0.065~~ - ~~\text{FL}; \\
&0.084+0.037~~ - ~~\text{ECH-M};\\
&0.074+0.025~~ - ~~\text{ECH-E-M};\\
&0.112+0.079~~ - ~~\text{IRR}.
\end{aligned}
\right.
\end{align}
\end{subequations}
For the pole mass of $b$-quark the procedure of  resummation of fermion loops, supplemented by the method of naive nonabelianization, predicts the decline of the high-order contributions up to six-loop level. A different picture is observed in estimates, made by other three methods: the asymptotic structure in expansion of pole mass of the bottom-quark manifests itself starting from the five-loop level. Another interesting feature is that within the ECH-motivated method which takes into account the transition from
the Euclidean to Minkowskian regions the PT series for pole mass of $b$-quark demonstrates output to some kind of plateau at four, five and six-loop levels. 
For the  case of the top-quark pole mass all four considered estimate procedures outline the decrease of the five and six-loop corrections. This means that the asymptotic structure of this PT series is not yet
manifesting itself at these levels. Therefore theoretical conception of pole mass of top-quark can be
safely used even at the six-loop level.
\section{Conclusion}
With help of the four methods, namely the resummation of the quark bubble chains, two ECH-motivated methods, defined in the Euclidean and Minkowskian regions, and the infrared renormalon-based approach, we estimate the five and six-loop corrections to the $\rm{\overline{MS}}$-on-shell heavy quark mass relation. In addition we determine flavour dependence of the considered contributions in the $\mathcal{O}(a^5_s)$ and $\mathcal{O}(a^6_s)$ orders. 
The IRR-based asymptotic  technique with $N_m$ normalization factor, taken in the four-loop approximation,   gives surprising non-oscillating in powers of  $n_l$ estimates,  while the both ECH and FL approaches predict not only close values of the corresponding coefficients but reproduce the sign-alternating structure of these corrections. The ECH Euclidean method for the  $b$-quark pole mass leads to the  effective plateau  and the rest three methods outline the increase of the five- and six-loop contributions. In the case of $t$-quark the asymptotic nature of the corresponding PT series is not observed even at six-loop level. Therefore theoretical concept of the pole mass of top-quark is applicable up to 6 order of PT for sure. Herewith the theoretical uncertainty of the existing values of the pole mass of the top quark is estimated by the last term of this asymptotic series, included in comparison with experimental data from the Tevatron and LHC colliders, which in the case of three-loop approximation is about 500 $\rm{MeV}$.
\section{Acknowledgments}
We would like to thank V.~M. Braun and A.~G. Grozin for helpful   discussions. The  work of V.~M. has been supported by the Russian Science Foundation grant No. 16-12-10151. 

\vspace{1cm}

\centering
\renewcommand\refname{
\end{document}